# Isotropic Negative Thermal Expansion Metamaterials

Lingling Wu,[1] Bo Li,[2] Ji Zhou[1]*

[1]State Key Laboratory of New Ceramics and Fine Processing, School of Materials Science and Engineering, Tsinghua University, Beijing 100084, China
[2]Advanced Materials Institute, Shenzhen Graduate School, Tsinghua University, Shenzhen, China

*To whom correspondence should be addressed. E-mail: zhouji@mail.tsinghua.edu.cn

**Abstract**:
Negative thermal expansion materials are important and desirable in science and engineering applications. However, natural materials with isotropic negative thermal expansion are rare and usually unsatisfied in performance. Here, we propose a novel method to achieve negative thermal expansion via a metamaterial approach. The metamaterial is constructed with unit cells that combine bi-material strips and anti-chiral structures. Both experimental and simulation results display isotropic negative thermal expansion properties. The coefficient of negative thermal expansion of the metamaterials is demonstrated to be dependent on the difference between the thermal expansion coefficients of two component materials in the bi-material strips, as well as on the circular node radius and the ligament length in the anti-chiral structures. The measured value of the linear negative thermal expansion coefficient reaches $-68.1 \times 10^{-6}$ 1/K in an operating temperature range from 303.15 K to 773.15 K, which is among the largest achieved in experiments to date. Our findings provide a novel and practical approach to obtaining materials with tunable isotropic negative thermal expansion on any scale.

**Main Text:**
Negative thermal expansion materials are highly desired in infrastructure engineering and precision instrument manufacturing[1-6]. However, few materials naturally exhibit negative thermal expansion characteristics because the phenomenon is derived from abnormal mechanisms such as phase transitions[7] and shortening of bond lengths[8], rigid unit modes[9], electronic effects[10], and magnetostriction[11]. Although several natural negative thermal expansion materials such as perovskite, $NaZn_{13}$-type $La(Fe,Si,Co)_{13}$ compounds[12], $Ag_3[Co(CN)_6]$[13], $(Hf, Mg)(WO_4)_3$[14] and the well-known $ZrW_2O_8$ family of materials[4,15] exist, few are widely used because of their narrow temperature range of negative thermal expansion, low thermal expansion coefficient value, anisotropy of thermal response[16-18] and low design freedom.

In the past decade, metamaterials have offered an entirely new route to constructing artificial materials with abnormal properties not present in the component materials. Several approaches have been proposed to achieve negative thermal expansion structures[19-28], however, because most of the negative thermal expansion cells in these metamaterials are asymmetric, i.e., their thermal expansion property is tailored in only one particular direction, few exhibit isotropic negative thermal expansion. In this paper, a novel model with isotropic negative thermal expansion is proposed and demonstrated through the use of anti-chiral structures composed of bi-metal strips. The metamaterials possess significant negative thermal expansion coefficients and broad operation temperature ranges. Both simulation and experimental results show that the structures exhibit isotropic negative thermal expansion and that tailored negative thermal expansion was achieved by changing the component materials and size parameters of the structures. Because the auxetic properties are independent of scale, the proposed structures can be applied at the macro-, micro- or even nano-scale[29].

The unit cell of the metamaterials is designed on the basis of anti-chiral structure—a well-established mechanical structure with an isotropic negative Poisson's ratio[30-32]. As proposed by Wojciechowski[33] and initially realized by Lakes and Sigmound[30,34], the anti-chiral structure is constructed as a central point (henceforth referred to as a "node") with several tangentially attached ligaments. In fact, the anti-chiral structures contain a series of similar systems when the number of attached ligaments (henceforth denoted as N) changes[35]. The unit cell is rotationally symmetrical of order $N$, where $N$ has been demonstrated to be restricted to only 3 (anti-trichiral) or 4 (anti-tetrachiral) in two-dimensional systems[36]. In a three-dimensional mechanism, only an anti-tetrachiral system can be constructed because a three-dimensional anti-trichiral structure cannot geometrically exist [37]. The existing patterns of anti-chiral systems are shown in Fig. 1(A).

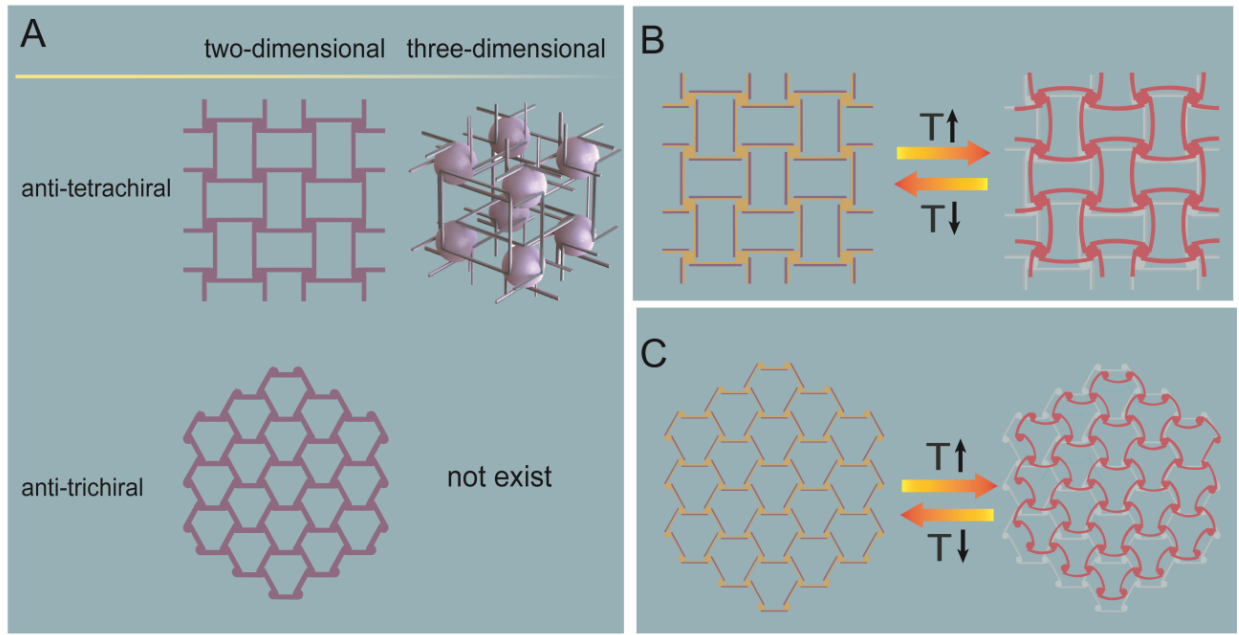

**Fig. 1.** (A) The existing two- and three-dimensional systems of anti-chiral metamaterials. (B) Deformation schematic of the anti-tetrachiral metamaterial under thermal stress. (C) Deformation schematic of the anti-trichiral metamaterial under thermal stress.

To obtain a negative thermal expansion effect, we introduced bi-material strips into the above anti-chiral structures. The linear coefficient of thermal expansion (CTE) $\alpha_L$ was used to define the thermal expansion of the model, which is calculated by the following equation:

$$\alpha_L = \frac{\Delta L}{\Delta T} \tag{1}$$

where $L$ is the length of the unit cell along the studied direction and $T$ is the temperature.

A bi-material structure composed of two materials with different CTEs has been demonstrated to bend with a curvature of $\frac{1}{\rho}$ under a thermal stress proportional to the difference of the CTEs of the component materials[38], as shown in Fig. 2A:

$$\frac{1}{\rho} \propto (\alpha_2 - \alpha_1) \qquad (2)$$

where

$\alpha_1$ and $\alpha_2$ are the CTEs of the component materials.

Combining the bi-material strips with the anti-chiral structures results in new negative thermal expansion metamaterials. Because equation (2) is known to be valid for bi-material strips made from thin ligaments where the end effects are neglected[39,40], both thermal stress and applied loading stress can cause material deformation. Therefore, the bending of ligaments in metamaterials due to thermal stress will result in a shortening of the node-node distance and a tendency to contract the structure in both in-plane principal directions, as shown in Figs. 1B and 1C. As a result, the entire system will exhibit isotropic negative thermal expansion.

To verify this hypothesis, theoretical simulation work was performed using the COMSOL Multiphysics commercial software. Metamaterials with two-dimensional anti-tetrachiral and anti-trichiral structures and a three-dimensional anti-tetrachiral structure constructed with two materials with different CTEs were considered. Schematics of the simulated two-dimensional anti-trichiral and anti-tetrachiral structures are shown in the inset of Fig. 2B.

We investigated the relationship between the effective linear CTE of the bi-material anti-chiral models and their constructed parameters, including the node radius ($r$), the length of ligaments ($l$) and the difference of CTE of component materials ($\alpha_2$ - $\alpha_1$). The detailed simulation setup is presented in the supporting online material (SOM). The simulated results are shown in Figs. 2B-D.

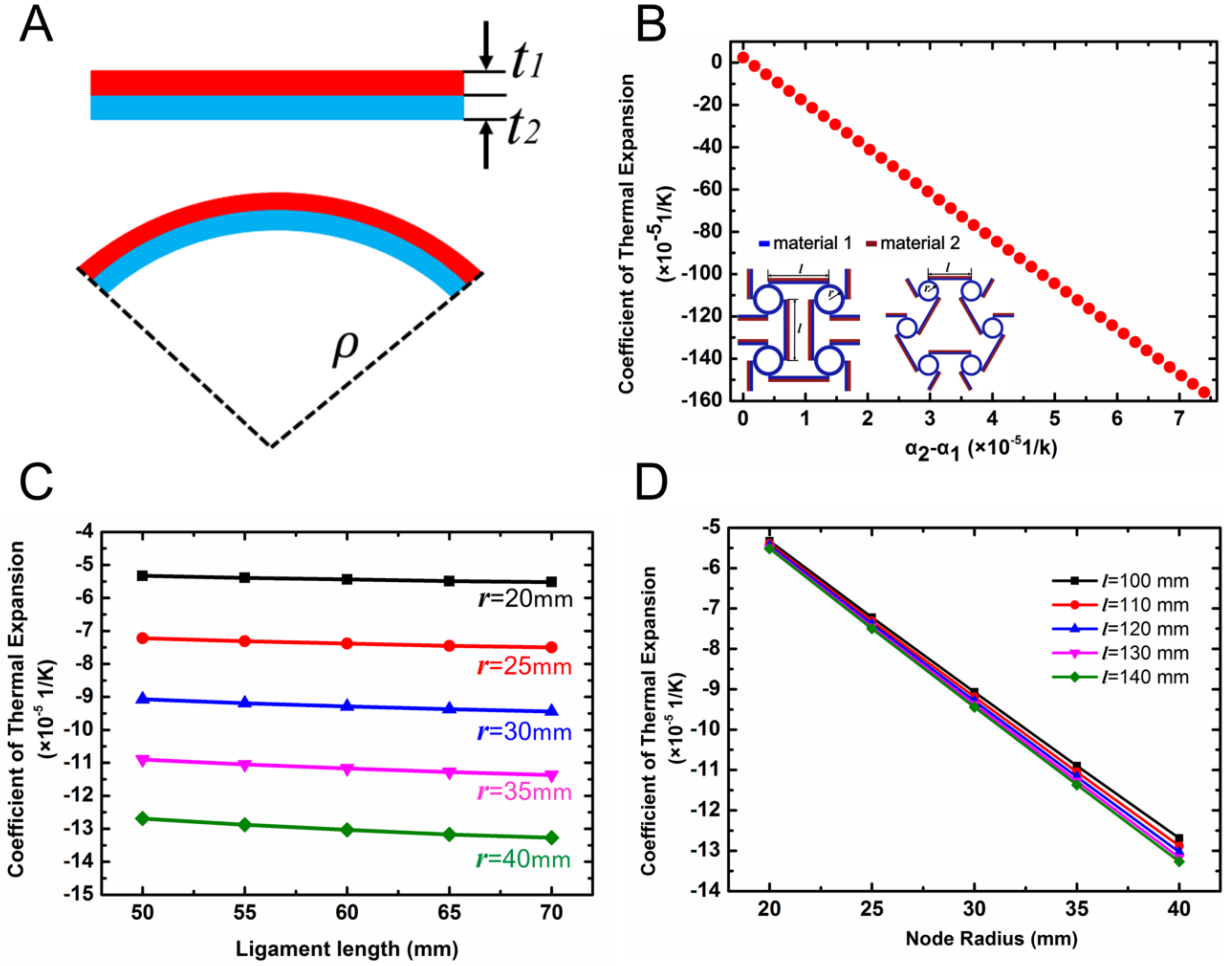

**Fig. 2.** (A) Schematic of the bi-material structure and its bending mechanism. (B) The relationship between the effective CTE of anti-chiral elements and the difference in CTE of their component materials. (C) The relationship between the ligament length ($l$) and the effective CTE of anti-chiral elements with different node radii ($r$). (D) The relationship between the circular node radius ($r$) and the effective CTE of anti-chiral elements with different lengths of ligament ($l$). In the simulation of (B), $l = 100$ mm, $r = 25$ mm, $\alpha_1 = 1.85 \times 10^{-5}$ 1/K, and $\alpha_2$ varies from $1.85 \times 10^{-5}$ to $9.25 \times 10^{-5}$ 1/K in increments of $1.85 \times 10^{-6}$ 1/K. In the simulation of (C) and (D), $r$ changes from 20 mm to 40 mm in increments of 5 mm and $l$ varies from 100 mm to 140 mm in increments of 10 mm. The inset of (B) shows the diagrams of the bi-material anti-tetrachiral element and the anti-trichiral element.

The results in Fig. 2B indicate that the effective CTE of the bi-material anti-chiral structures is proportional to the difference between the CTEs of the component materials. Furthermore, the relationship curves for the anti-trichiral and anti-tetrachiral structures are exactly the same when the circular node radius ($r$) and ligament length ($l$) are constant. The simulated relationship curves between the effective CTEs of the models and their size parameters are plotted in Figs. 2C and 2D, which show that, for both the anti-trichiral and anti-tetrachiral structures, the effective CTE is proportional to the node radius $r$ and ligament length $l$. Through mathematical analysis, we concluded that the relationship between the effective CTE of the anti-chiral models and their constructed parameters, including the difference of CTE in the component materials ($\alpha_2 - \alpha_1$), circular node radius $r$ and the ligament length $l$, is described by the following empirical equation:

$$\alpha_{eff} \propto (\alpha_2 - \alpha_1) \bullet (l - A) \bullet (r - B) \qquad (3)$$

where $\alpha_{eff}$ is the effective CTE of the anti-chiral models and $A$ and $B$ are constants.

Equation (3) shows that tunable negative thermal expansion can be achieved by carefully selecting the component materials and adjusting the geometrical parameters of the bi-material anti-chiral models. Typical simulated deformation patterns of two-dimensional anti-tetrachiral and anti-trichiral structures with a node radius $r$ of 40 mm and a ligament length $l$ of 100 mm are shown in Figs. 3A and 3D, and the complete results are presented in the SOM (Fig. S1). As evident in these figures, when the temperature was increased from 303.15 K to 773.15 K, both structures shrank isotopically.

To experimentally verify the negative thermal expansion property of the metamaterials, we fabricated samples of bi-metal anti-chiral structures with a ligament length $l$ of 100 mm and a variable node radius $r$. The samples were constructed with aluminum and copper as the higher and lower CTE component materials, respectively. The fabricated parts of the anti-tetrachiral samples are shown in Fig. S5C. Details of the fabrication process are presented in the SOM (Figs. S2 and S5C).

The experimental conditions were set exactly the same as those in the simulation. The temperature was uniformly increased from 303.15 K to 773.15 K by a heating platform. The detailed measurement setup is presented in the SOM (Fig. S5). To accurately measure the deformation of the samples, we used commercial measurement software (MindVision) to determine the sizes of the samples at 303.15 K and 773.15 K. The measurement results for anti-tetrachiral and anti-trichiral samples with a node radius of 40 mm are shown in Figs. 3B-C and Figs. 3E-F. The samples in Figures 3B and 3E were measured at 303.15 K, and those in Figs. 3C and 3F were measured at 773.15 K. To improve the measurement accuracy, we made two measurements along each orthogonal direction for the two-dimensional anti-tetrachiral samples and calculated the average values. The same process was used for the anti-trichiral samples, except that we measured three lengths along three directions with equal included angles because the anti-trichiral model exhibits threefold rotational symmetry. The complete measurement results for the two structures are shown in the SOM (Figs. S3 and S4).

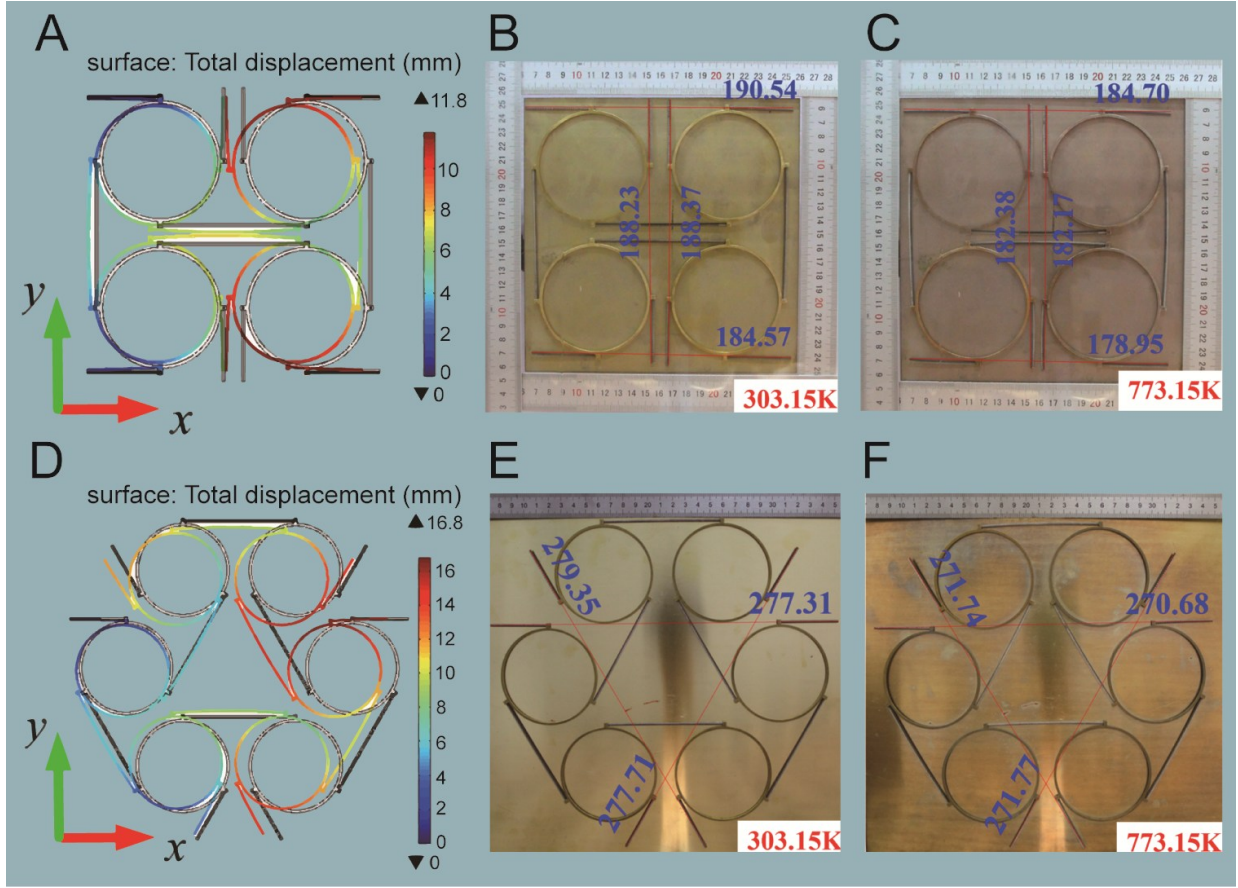

**Fig. 3. (A)** The simulated deformation of a two-dimensional anti-tetrachiral system with $r = 40$ mm and $l = 100$ mm. **(B)(C)** The experimentally measured results of an anti-tetrachiral sample with $r = 40$ mm and $l = 100$ mm at 303.15 K and 773.15 K, respectively. **(D)** The simulated deformation of a two-dimensional anti-trichiral structure with $r = 40$ mm and $l = 100$ mm. **(E)(F)** The experimentally measured results of an anti-trichiral sample with $r = 40$ mm and $l = 100$ mm at 303.15 K and 773.15 K, respectively. For both the simulation and experiments, $\alpha_1 = 1.85 \times 10^{-5}$ 1/K and $\alpha_2 = 2.3 \times 10^{-5}$ 1/K. The measurement unit in (B)(C) and (E)(F) is millimeters.

The experimentally measured relationship between the effective CTE and the node radius ($r$) for the anti-tetrachiral and anti-trichiral samples are plotted in Figs. 4A-C to provide a better comparison with the simulation results. As evident in these figures, both the simulation and experimental results show that the structures exhibit isotropic negative thermal expansion and the values of effective CTEs are proportional to the node radius ($r$). Reasonable agreement was achieved between the simulation and experimental results, with the experimental data generally showing lower values of the thermal expansion coefficient compared to the simulation results. This discrepancy is likely due to the oxidation of bi-metal strips in the air and measurement errors. In Figs. 4A and 4B, the measured value of the effective CTE for the anti-tetrachiral system reaches $-68.1 \times 10^{-6}$ 1/K over a large operating temperature range from 303.15 K to 773.15 K; this isotropic negative CTE value is among the largest reported to date. Compared with existing artificial negative thermal expansion materials, our models are easier to fabricate, exhibit better performance and have greater application potential in engineering fields.

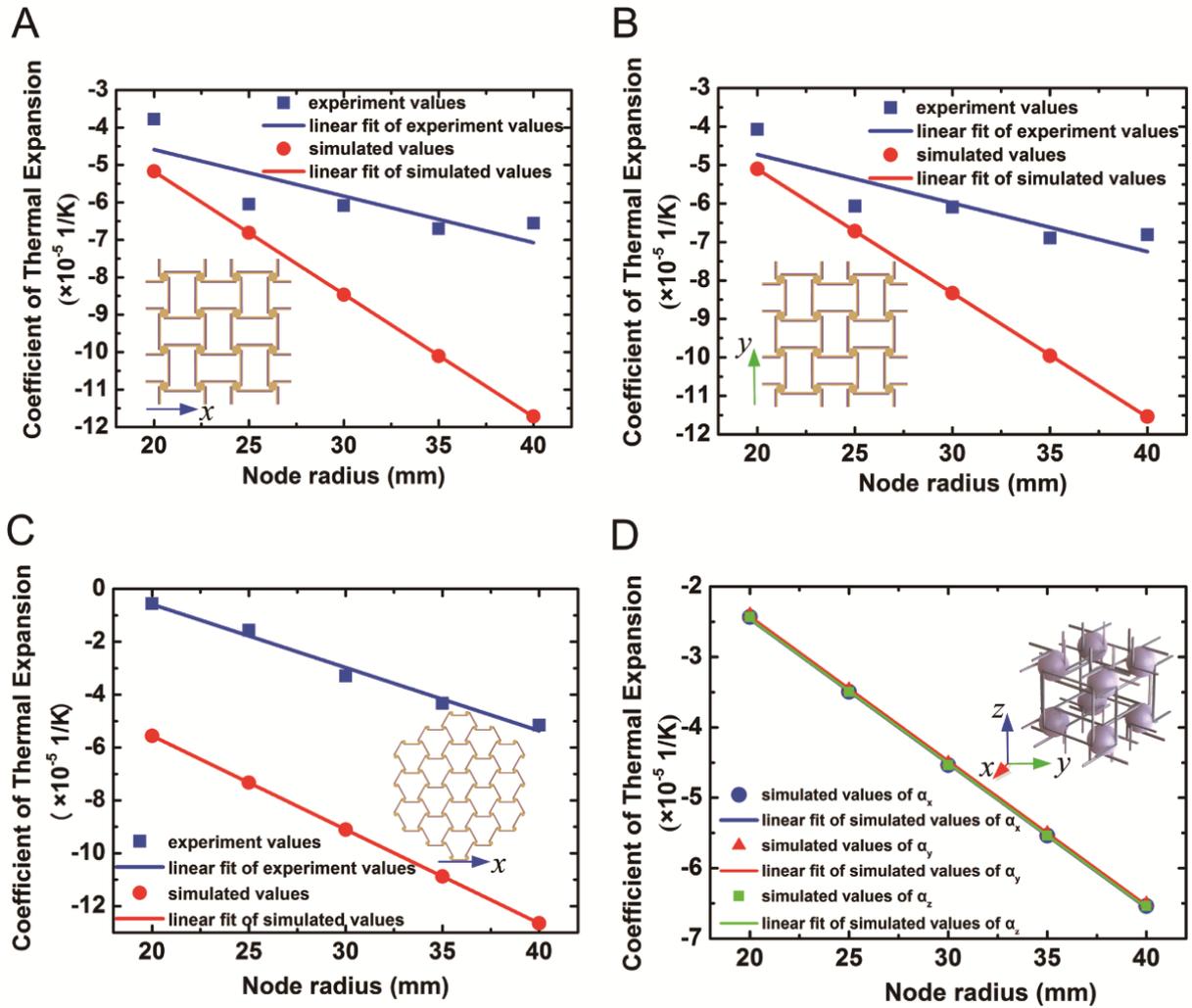

**Fig. 4.** **(A)(B)** Comparison of the simulated and experimentally measured relationship between the node radius ($r$) and the effective CTEs along two orthogonal directions for the two-dimensional anti-tetrachiral samples. **(C)** Comparison of simulated and experimentally measured relationship between the node radius ($r$) and the effective CTE along $x$ directions for the two-dimensional anti-trichiral samples. **(D)** The simulated relationship between the node radius ($r$) and the effective CTEs of the three-dimensional anti-tetrachiral systems along three orthogonal directions. The insets are models of each corresponding structure.

In conclusion, we proposed three- and two-dimensional metamaterials with substantial isotropic negative thermal expansion capabilities and demonstrated their fabrication by combining anti-chiral structures and bi-material strips. The relationship between the negative thermal expansion properties and the physical parameters of the composed materials and the structural parameters of the metamaterials were established. Large and tailorable isotropic negative CTEs were obtained with this approach. The mechanism proposed here is theoretically scale independent, meaning that the concept of metamaterial manipulation of thermal expansion could be extended to the micro- or nano-scale if appropriate component materials and fabrication processes are available.

These metamaterials have extensive potential applications in areas such as aerospace engineering, mechanical engineering, and architectural engineering.

**Methods**
  **Theory and simulation**

A metamaterial approach is proposed to achieve tunable isotropic thermal expansion by combining a bi-material structure and anti-chiral systems.

A bi-material structure composed of two materials with different thermal expansion coefficients (CTE) has been demonstrated to bend with a curvature of $1/\rho$ [38]:

$$\frac{1}{\rho} = \frac{6(1+m)^2 (\alpha_2 - \alpha_1)\Delta T}{(t_1 + t_2)\left[3(1+m)^2 + (1+mn)(m^2 + \frac{1}{mn})\right]} \quad (S1)$$

As shown in Fig. 2A, if we note $\frac{t_1}{t_2} = m$, $\frac{E_1}{E_2} = n$,

Where
$\rho$ is the radius of curvature of the bi-material strip;
$t_1$ and $t_2$ are the thickness of two components of the bi-material strip;
$E_1$ and $E_2$ are the Young's modulus of each material;
$\alpha_1$ and $\alpha_2$ are the thermal expansion coefficient of the component materials;
$\Delta T$ is the temperature variation.

Equation (S1) is known to be valid for bi-material strips fabricated from thin ligaments where end effects are neglected [39,40]. On the basis of equation (S1), the curvature of the bi-material strip is proportional to the difference of CTE of the component materials when other parameters are constant. To simplify the model, we set $t_1 = t_2 = t$, i.e., $m = 1$, and equation (S1) becomes

$$\frac{1}{\rho} = \frac{12(\alpha_2 - \alpha_1)\Delta T}{t(14 + n + \frac{1}{n})} \quad (S2)$$

Several authors have reported the auxetic property of anti-chiral structures[29,30,33-35] that expand laterally when a uniaxial load is applied. Therefore, we hypothesize that, if a bi-material structure and an anti-chiral structure are combined, the entire system may exhibit negative thermal expansion.

To verify this hypothesis and investigate the thermal performance of bi-material anti-chiral structures, we initially carried out simulation work using the COMSOL Multiphysics commercial software. To simplify the simulation work, the thickness of both component materials were set to 1 mm, the ligament length (*l*) was varied from 100 mm to 140 mm in increments of 10 mm, and the node radius (*r*) was varied from 20 mm to 40 mm in increments of 5 mm. To study the relationship between the effective CTE of the elements and the CTE of the component materials, we set the CTE of material 1 ($\alpha_1$) to $1.85 \times 10^{-5}$ 1/K and increased the CTE of material 2 ($\alpha_2$) from $1.85 \times 10^{-5}$ 1/K to $9.25 \times 10^{-5}$ 1/K in increments of $1.85 \times 10^{-6}$ 1/K. The simulated temperature range was 303.15 K to 773.15 K. The two-dimensional simulation models are shown in the inset of Fig. 2B. The simulated relationship curves between the effective CTE and the difference of the component material CTEs ($\alpha_2 - \alpha_1$), node radius ($r$) and ligament length ($l$) are shown in Figs. 2B-D, respectively.

For comparison with the experimental work, simulations were also performed for both the two-dimensional anti-trichiral and anti-tetrachiral structures with $l = 100$ mm, $\alpha_1 = 1.85 \times 10^{-5}$ 1/K, $\alpha_2 = 2.3 \times 10^{-5}$ 1/K and a variable node radius ($r$). Because a three-dimensional anti-trichiral structure could not be constructed[37], a three-dimensional anti-tetrachiral structure was investigated in the subsequent simulation. The simulated deformation patterns are shown in Fig. S1, which shows that both the two-dimensional anti-tetrachiral and anti-trichiral structures exhibit negative thermal expansion along two orthogonal directions. Figs. S1 K-O show that the three-dimensional anti-tetrachiral structure also exhibits negative thermal expansion along three orthogonal directions.

**Fabrication process**

The model requires two component materials with different thermal expansion coefficients. We selected a bi-metal plate with aluminum and copper as the raw materials. Aluminum has a relatively high thermal expansion coefficient of $2.3 \times 10^{-5}$ 1/K, whereas copper has a lower coefficient of $1.85 \times 10^{-5}$ 1/K. A cutting machine was used to cut the bi-metal plates into strips with a thickness of 1 mm. The circular node was designed as a ring with grooved ears, and both the grooved ears and bi-metal strips were drilled so that they could be firmly tightened with a screw, as shown in Fig. S5C. The fabricated two-dimensional anti-chiral elements with node radii that vary from 20 mm to 40 mm are shown in Fig. S2.

**Testing setup**

A measurement setup was constructed as shown in Fig. S5 to verify the thermal expansion properties of the proposed models. The experimental setup was composed of three parts: the heating platform and its controller, the CCD camera and a computer. The sample was placed on the heating platform and covered with a quartz jar to prevent air convection, as shown in Fig. S5A. The computer-controlled CCD camera was positioned above the heating platform to observe the deformation of the sample in real time. To ensure soft lighting during the experiment, two light-compensating lamps were placed above the measurement equipment.

**Acknowledgments:** The authors gratefully acknowledge financial support by fund from the National Natural Science Foundation of China (Nos. 51032003, 11274198, 51221291) and the Science and Technology Plan of Beijing (Z141100004214001).

**Author contributions**

Ji Zhou conceived the idea. Lingling Wu designed experiments and carried out numerical calculations. Bo Li revised the full paper. All authors contributed to scientific discussion and critical revision of the article.

**Additional information:**

Competing financial interests: The authors declare no competing financial interests.

**Supporting Online Material：**

Figs. S1 to S5

# Supplementary Materials for

## Isotropic Negative Thermal Expansion Metamaterials

Lingling Wu, Bo Li, Ji Zhou*
*correspondence to: zhouji@mail.tsinghua.edu.cn

**This PDF file includes:**

Figs. S1 to S5

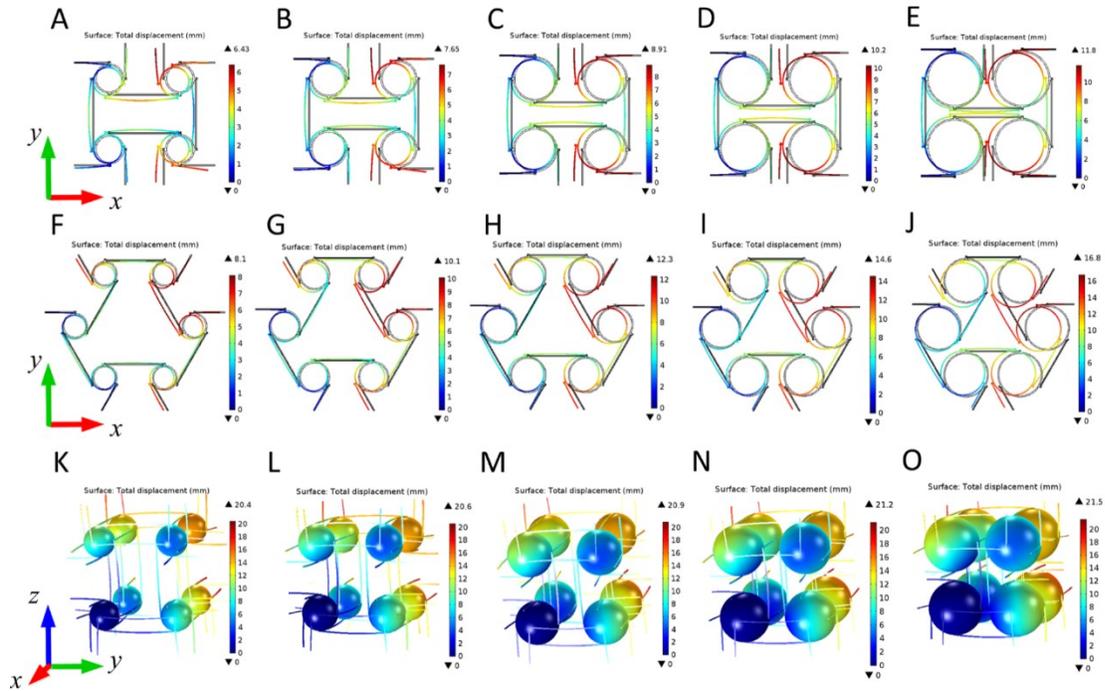

**Fig. S1**

Simulation results for the anti-chiral structures in COMSOL Multiphysics. (A-E) Simulation results for two-dimensional anti-tetrachiral structures with node radii of (A) 20 mm; (B) 25 mm; (C) 30 mm; (D) 35 mm; and (E) 40 mm. (F-J) Simulation results for two-dimensional anti-trichiral structures with node radii of (F) 20 mm; (G) 25 mm; (H) 30 mm; (I) 35 mm; and (J) 40 mm. (K-O) Simulation results for three-dimensional anti-tetrachiral structures with node radii of (K) 20 mm; (L) 25 mm; (M) 30 mm; (N) 35 mm; and (O) 40 mm.

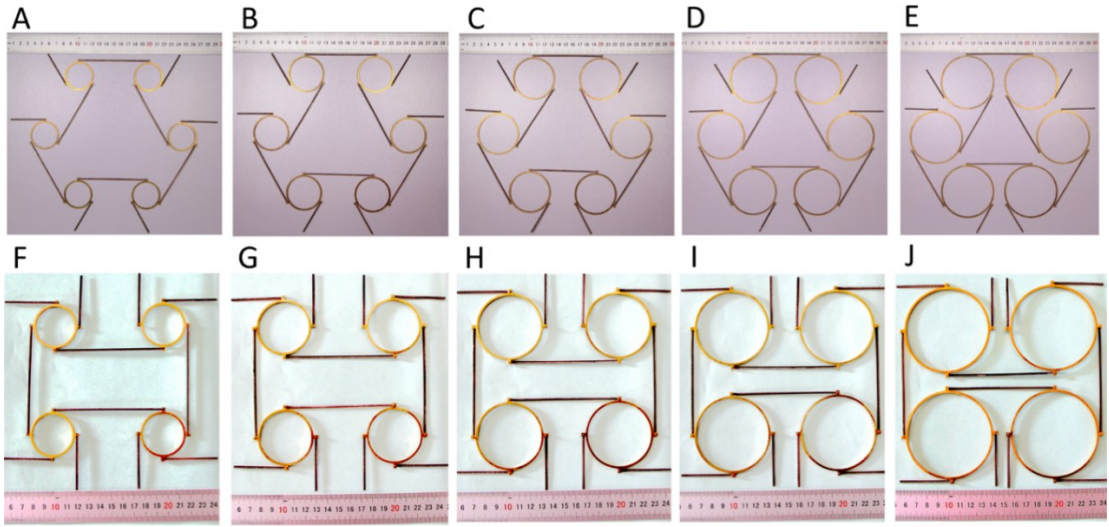

**Fig. S2**

Samples fabricated for two-dimensional anti-chiral structures: (A-E) Anti-trichiral structures with node radii of (A) 20 mm; (B) 25 mm; (C) 30 mm; (D) 35 mm; and (E) 40 mm. (F-J) Anti-tetrachiral elements with node radii of (F) 20 mm; (G) 25 mm; (H) 30 mm; (I) 35 mm; and (J) 40 mm.

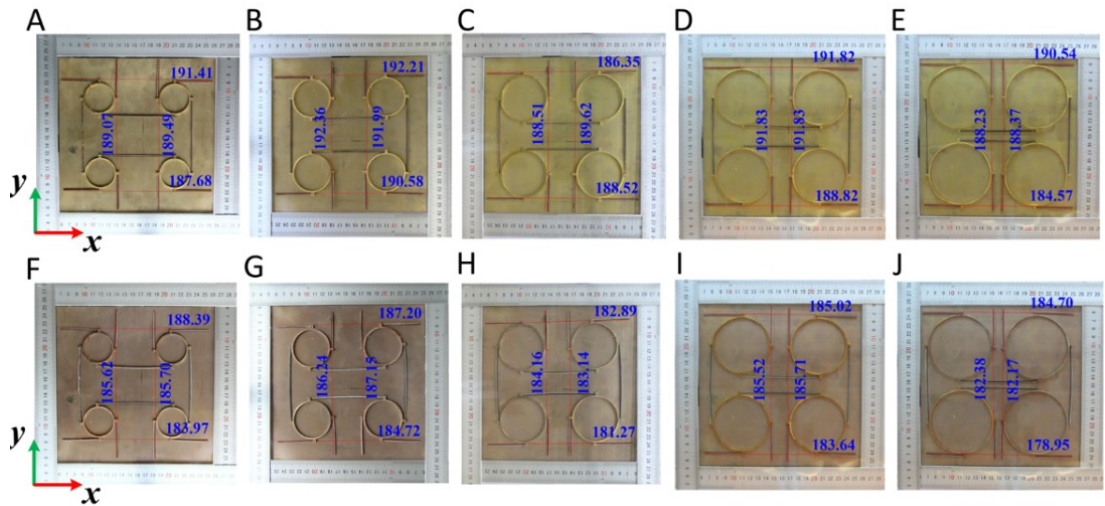

**Fig. S3**

Experimental results for two-dimensional anti-tetrachiral elements. (A-E) The measured size of samples with node radii of (A) 20 mm; (B) 25 mm; (C) 30 mm; (D) 35 mm; and (E) 40 mm at 303.15 K. (F-J) The measured size of samples with node radii of (F) 20 mm; (G) 25 mm; (H) 30 mm; (I) 35 mm; and (J) 40 mm at 773.15 K.

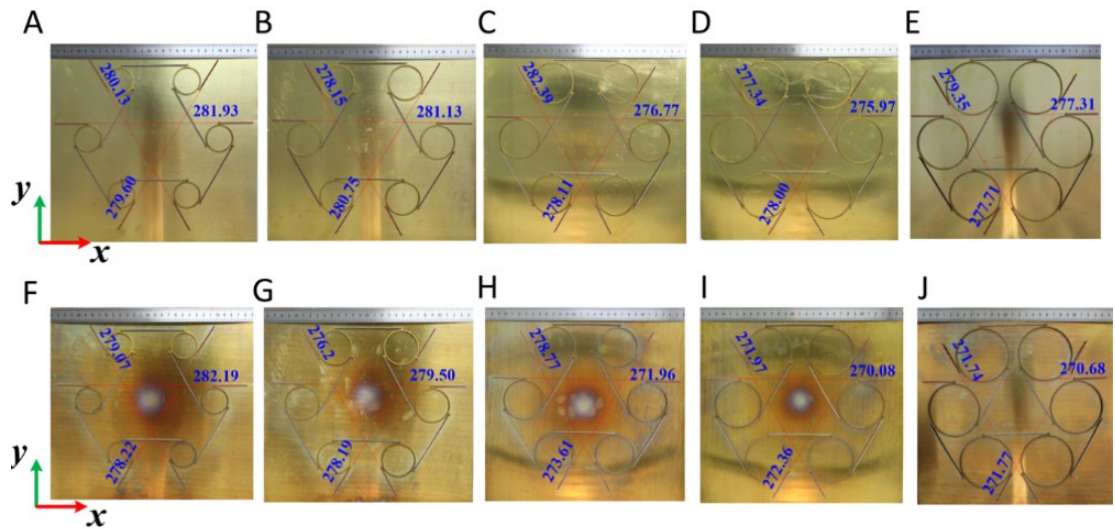

**Fig. S4**

Experimental results for two-dimensional anti-trichiral elements with different node radii. (A-E) The measured size of samples with node radii of (A) 20 mm; (B) 25 mm; (C) 30 mm; (D) 35 mm; and (E) 40 mm at 303.15 K. (F-J) The measured size of samples with node radii of (F) 20 mm; (G) 25 mm; (H) 30 mm; (I) 35 mm; and (J) 40 mm at 773.15 K.

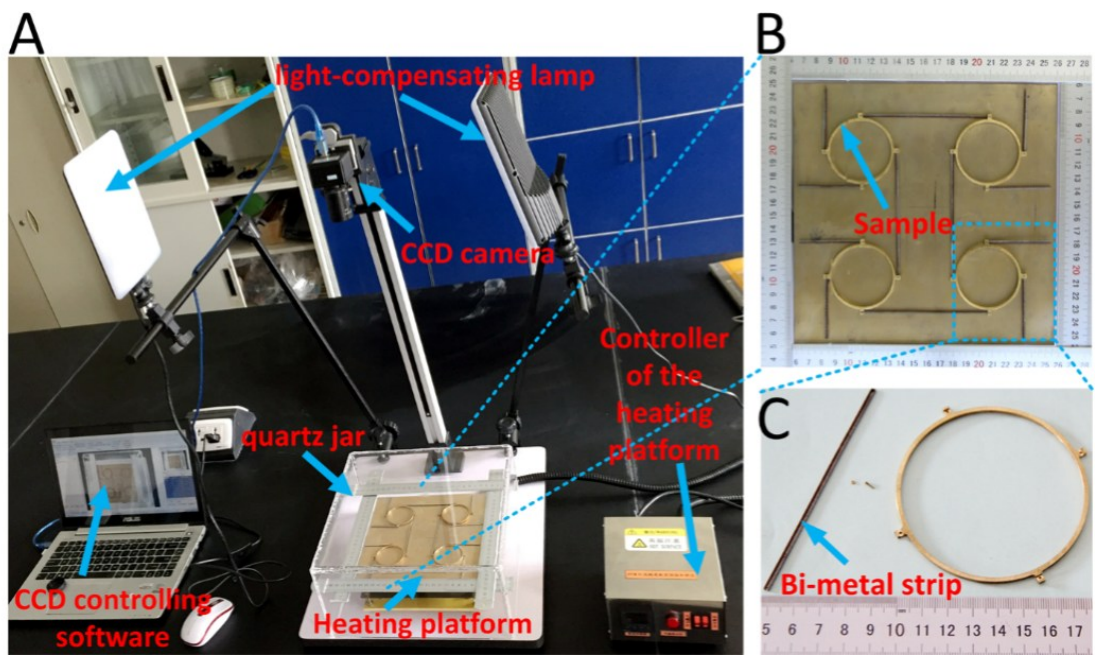

**Fig. S5**

Measurement setup. (A) View of the measurement setup. (B) The fabricated anti-tetrachiral sample placed on the heating platform. (C) Fabricated parts for the two-dimensional anti-chiral samples.